\documentclass[reprint,a4paper, amsfonts, amssymb, amsmath, reprint, showkeys, nofootinbib, twoside, superscriptaddress]{revtex4-1}
\usepackage[english]{babel}
\usepackage[utf8]{inputenc}
\usepackage[colorinlistoftodos, color=green!40, prependcaption]{todonotes}
\usepackage{amsthm}
\usepackage{mathtools}
\usepackage{physics}
\usepackage{xcolor}
\usepackage{graphicx}
\usepackage[left=23mm,right=13mm,top=35mm,columnsep=15pt]{geometry} 
\usepackage{adjustbox}
\usepackage{placeins}
\usepackage[T1]{fontenc}
\usepackage{lipsum}
\usepackage{csquotes}

\input{Commandes.sty}
\usepackage{amsmath}
\usepackage{amssymb}
\usepackage{amsfonts}
\usepackage{mathtools}
\usepackage{booktabs}
\usepackage{stmaryrd}
\usepackage{mathrsfs}
\usepackage{multirow}
\usepackage{centernot}

\usepackage{fancyvrb}
\usepackage{graphicx}
\usepackage{floatrow}
\usepackage{subfig}
\usepackage{wrapfig}
\usepackage{cases}
\usepackage{tikz}
\usepackage{eurosym} \def\{\euro{}} 
\usepackage{cmlgc}
\usepackage{color}
\usepackage[squaren,Gray]{SIunits}
\usepackage{empheq}
\usepackage[most]{tcolorbox}
\usepackage{booktabs}
\usepackage{textcomp} 
\usepackage{soul}
\usepackage{natbib}
\usepackage{filecontents}

\usepackage{graphicx}
\usepackage{txfonts}

\usepackage{url}
\definecolor{linkcolor}{rgb}{0,0,0}
\definecolor{linkcolorurl}{rgb}{0,0,1}
\usepackage[pdftex,colorlinks=true,
pdfstartview=FitV,
linkcolor= blue,
citecolor= blue,
urlcolor= blue,
hyperindex=true,
hyperfigures=true]
{hyperref}
\hypersetup{linktoc=page}

\begin{document}
\title{Multi-dimensional simulations of ergospheric pair discharges around black holes}

\defcitealias{Parfrey_2019}{P19}

\author{Benjamin Crinquand}
    \email[Correspondence email address: ]{benjamin.crinquand@univ-grenoble-alpes.fr}
    \affiliation{Univ. Grenoble Alpes, CNRS, IPAG, 38000 Grenoble, France\\}
\author{Benoît Cerutti}
    \affiliation{Univ. Grenoble Alpes, CNRS, IPAG, 38000 Grenoble, France\\}
\author{Alexander Philippov}
    \affiliation{Center for Computational Astrophysics, Flatiron Institute, 162 Fifth Avenue, New York, NY 10010, USA}
\author{Kyle Parfrey}
    \affiliation{Department of Astrophysical Sciences, Peyton Hall, Princeton University, Princeton, NJ 08544, USA}
\author{Guillaume Dubus}
    \affiliation{Univ. Grenoble Alpes, CNRS, IPAG, 38000 Grenoble, France\\}

\date{\today} 

\begin{abstract}
Black holes are known to launch powerful relativistic jets and emit highly variable gamma radiation. How these jets are loaded with plasma remains poorly understood. Spark gaps are thought to drive particle acceleration and pair creation in the black-hole magnetosphere. In this paper, we perform 2D axisymmetric general-relativistic particle-in-cell simulations of a monopole black-hole magnetosphere with a realistic treatment of inverse Compton scattering and pair production. We find that the magnetosphere can self-consistently fill itself with plasma and activate the Blandford-Znajek mechanism. A highly time-dependent spark gap opens near the inner light surface which injects pair plasma into the magnetosphere. These results may account for the high-energy activity observed in active galactic nuclei and explain the origin of plasma at the base of the jet.
\end{abstract}



\maketitle

Active galactic nuclei (AGN) can be responsible for the launching of powerful relativistic plasma jets. Very long baseline interferometry shows that these jets are launched very close to the event horizon of the black hole~\citep{Walker_2018}, implying that processes occurring in its close environment must be at play. Some AGN are also known to emit ultra-rapid gamma-ray flares \citep{Abramowski_2012, Aharonian_2006} suggesting that sub-horizon scales, possibly at the base of the jet, are involved in efficient particle acceleration. Non-thermal emission from accelerated particles was recently detected in the immediate vicinity of the AGN M87* \citep{EHT_1}. This creates new opportunities to better understand black-hole activity, as the black-hole system can now be directly probed down to sub-horizon scales.

 
A possible explanation for jet launching is provided by the Blandford-Znajek (BZ) mechanism~\citep{Blandford_1977}, which involves a force-free magnetosphere coupled to the black hole. This mechanism requires plasma to be continuously replenished, in order to sustain the force-free magnetosphere and to carry the Poynting flux. The jet generally comprises the magnetic field lines which enter the ergosphere and cross the event horizon. Since these field lines are disconnected from the disk, it is very unlikely that plasma from the accretion flow can fill the jet zone.

As the plasma density drops, the electric field induced by the rotation of the black hole becomes unscreened, leading to electrostatic gaps and particle acceleration. High-energy emission may result from inverse Compton (IC) scattering of soft photons by ultra-relativistic leptons. In this framework, annihilation between the high-energy photons produced in the gap and soft photons emitted by the accretion flow is a possible plasma source~\citep{Levinson_2011}. Electrostatic gaps could then both explain the observed gamma-ray flares and provide pair plasma to the jet. 


There have been numerous attempts to derive analytically the properties of a steady gap~\citep{Broderick_2015,Hirotani2016}, but the spark gap dynamics are most likely intermittent \citep{Levinson_2017}. The exact location of the gap is also unknown. The validity of the BZ mechanism has been demonstrated by general relativistic magnetohydrodynamic (GRMHD) simulations~\citep[e.g.][]{Komissarov_2004b}, but this numerical approach cannot address the questions of the source of plasma or particle acceleration. Kinetic simulations, on the other hand, can capture these effects. 1D general relativistic particle-in-cell (GRPIC) simulations display a time-dependent gap~\citep{Levinson_2018,Chen_2019}. \citet[][hereafter P19]{Parfrey_2019} performed the first global 2D GRPIC simulations of a nearly force-free magnetosphere. They ignored radiative transfer and instead injected pairs in proportion to the local parallel electric field. This prescription mimics pair creation, but precludes any chance of seeing a gap develop. 

In this work, we present 2D global GRPIC simulations with self-consistent radiative transfer, in order to model realistic plasma injection and study the spark gap dynamics. Both IC scattering and $\gamma\gamma$ pair creation processes are implemented. We use a general relativistic version of the PIC code \texttt{Zeltron}~\citep{Zeltron,Cerutti_2013,Cerutti_2015}, first introduced in \citetalias{Parfrey_2019}. The background space-time is described by the Kerr metric, with dimensionless spin parameter $a \in [0,1[$. We use Kerr-Schild spherical coordinates $(t,r,\theta,\varphi)$, which do not possess a coordinate singularity at the event horizon. For convenience, we define ``fiducial observers'' (FIDOs), whose wordlines are orthogonal to spatial hypersurfaces.

We include gamma-ray photons in our simulations as a neutral third species that follows null geodesics. We extended to full 3D the radiative transfer algorithm of \citet{Levinson_2018}, which incorporates IC scattering and photon-photon pair production (see the Supplemental Material, which includes Refs.~\citep{Novikov,Blumenthal_1970,Jones_1968,Gould_1967,Bonometto_1971,Aharonian_1983}). Electrons, positrons and gamma-ray photons interact with a background  radiation field of soft photons. For simplicity, we assume that the radiation field is time-independent, uniform, isotropic and mono-energetic, with energy $\varepsilon_0$ and density $n_\mathrm{0}$. We do not include any feedback of the simulation on this radiation field. The upscattered photons and created leptons are assumed to propagate along the same direction as their high-energy parents, reflecting strong relativistic beaming. The fiducial optical depth of both processes is $\tau_0=n_\mathrm{0} \sigma_T r_g$, where $r_g$ is the gravitational radius and $\sigma_T$ is the Thomson cross-section.

\begin{figure*}[ht!]
    \centering
    \resizebox{\hsize}{!}{\includegraphics{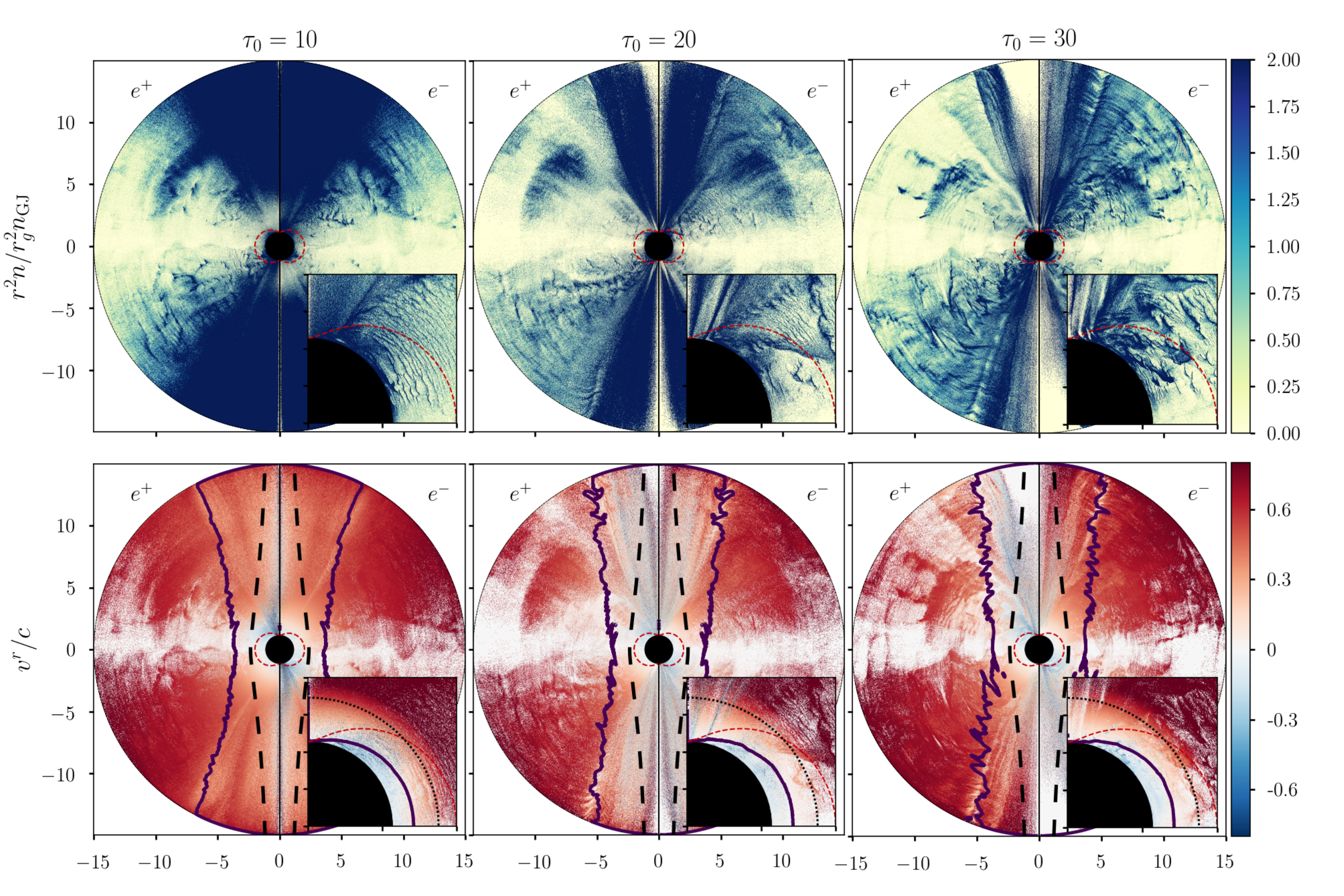}}
	\caption{Top panel: snapshots of the steady-state normalized densities $n$ for positrons (left) and electrons (right), compensated by $r^2$, for three fiducial optical depths $\tau_0=10$, $20$ and $30$. Insets show the density close to the horizon. Bottom panel: snapshots of the steady-state radial $3$-velocities $v^r$ for positrons (left) and electrons (right) for $\tau_0=10$, $20$ and $30$. The loosely dashed black line is the stagnation surface given by \citep{Hirotani2016}. Insets show the $3$-velocity close to the black hole, with higher contrast to help visualize the change in sign. The two solid lines are the inner and outer light surfaces. The dotted black line is the null surface as given by \citep{Levinson_2017}. In all plots, the densely dashed red line marks the ergosphere. All distances are in units of $r_g$.}
    \label{fig:densities}
\end{figure*}

In this paper we choose to endow the black hole with a monopole magnetic field (see the Supplemental Material). Although unphysical, this magnetic configuration has several benefits. (i) Our results can be directly compared to the BZ analytical solution, which assumes a magnetic monopole. (ii) We can capture the intrinsic physical properties of the gap without interference from more complex structures, such as current sheets. (iii) It is a realistic model for the field lines penetrating the ergosphere on each hemisphere, irrespective of the magnetosphere's large-scale structure~\cite{Komissarov_2004b,Komissarov_2007}.

We use a 2D axisymmetric setup with spherical coordinates $(r,\theta)$. The simulation domain is $r\in [r_\mathrm{min}=0.9\, r_h, r_\mathrm{max}=15 \, r_g]$, $\theta \in [0,\pi]$, where $r_h = r_g (1+\sqrt{1-a^2})$ is the radius of the event horizon. The ergosphere is the region within the axisymmetric surface defined by $r=r_g(1+\sqrt{1-a^2 \cos^2{\theta}})$. The spin parameter is set at $a=0.99$. The spatial grid points are uniformly spaced in $\log_{10} r$ and $\theta$. We mimic an open outer boundary using an absorbing boundary layer~\citep{Cerutti_2015}. Particles are removed if $r\le r_h$ or $r\ge r_\mathrm{max}$. We performed our runs with a grid resolution $2000 \, (r) \, \times \, 1152 \, (\theta)$, with the requirement that we resolve the plasma skin depth everywhere. This was checked \emph{a posteriori} since the plasma density is one of the unknowns. Initially, the magnetosphere is empty of pairs but filled with gamma-ray photons distributed uniformly and isotropically from $r=r_h$ to $r=4 r_h$, with the energy $\varepsilon_1=400 m_e c^2$, which is well above the pair creation threshold. The photons quickly pair produce, igniting the pair discharge.

We use normalized code units where $r_g$ is the unit of length and $r_g/c$ the unit of time. The normalized magnetic field is $\tilde{B}_0 = r_g (e B_0 / m_e c^2)$, and the normalized energy of background photons is $\tilde{\varepsilon}_0 = \varepsilon_0 / m_e c^2$. Three dimensionless parameters define the physical conditions around the black hole: $\tilde{B}_0$, $\tilde{\varepsilon}_0$ and $\tau_0$. In M87*, the magnetic field is estimated to be $B_0 \approx 100$ G ($\tilde{B}_0 \sim 10^{14}$) \citep{Neronov_2007,EHT_5}, whereas the soft background photon field peaks at $\varepsilon_0 \approx 1$ meV ($\tilde{\varepsilon}_0 \sim 10^{-9}$) \citep{Abdo_2009}. The optical depth is uncertain, but is likely to be $\lesssim 10^{3}$ \citep{Levinson_2011,Moscibrodzka_2011}. The density scale needed to screen the vacuum parallel electric field is the typical Goldreich-Julian number density $n_\mathrm{GJ} =B_0 \omega_\mathrm{BH} /(4 \pi c e )$~\citep{Goldreich_1969}, taking $\omega_\mathrm{BH}=ca/(2r_h)$ as the black-hole angular velocity. 


The maximum Lorentz factor $\gamma_\mathrm{max}$ that leptons can reach is close to $a \tilde{B}_0$. We also define $\gamma_s$ as the typical Lorentz factor of secondary particles that have just been pair produced. We focus our work on AGN characterized by $1~\ll~\gamma_s~\ll~\gamma_\mathrm{max}$. The cross-section of $\gamma\gamma$ pair production peaks near the threshold \citep{Gould_1967}, so the bulk of pairs is created at $\gamma_s \sim 1/\tilde{\varepsilon}_0$. The greater the ratio $\gamma_\mathrm{max}/ \gamma_s \sim a \tilde{B}_0 \tilde{\varepsilon}_0$, the higher the resulting plasma multiplicity (defined as the plasma density normalized by $n_\mathrm{GJ}$) will be~\citep{Timokhin_2019}. 

Altogether, we must choose $\tilde{\varepsilon}_0$ low enough, so that $\gamma_s~\gg~1$, but $\tilde{B}_0 \tilde{\varepsilon}_0$ large enough, to guarantee a good separation of scales and a large multiplicity ($\gamma_s \ll \gamma_\mathrm{max}$). In practice we chose $\tilde{B}_0=5\times 10^5$ and $\tilde{\varepsilon}_0=5\times10^{-3}$. The product $\tilde{B}_0 \tilde{\varepsilon}_0 =2500$ is still two orders of magnitude below its estimated value for M87*, but it is large enough to induce a transition to time-dependent gaps at high opacity. We checked that with these parameters, particle acceleration is not limited by radiative IC losses. For lower values of $\tilde{B}_0 \tilde{\varepsilon}_0$, the gaps remain steady at all $\tau_0$. On the other hand, increasing the magnetic field implies decreasing the plasma skin depth $d_e = \sqrt{m_e c^2 / 4 \pi n_\mathrm{GJ} e^2} \sim r_g \tilde{B_0}^{-1/2}$, so the resolution needs to go up. We are thus limited to unrealistically low values for $\tilde{B}_0$ and high values for 
$\tilde{\varepsilon}_0$, since $\tilde{B}_0 \tilde{\varepsilon}_0$ must remain large.

\begin{figure}[b!]
    \centering
    \resizebox{1.0\hsize}{!}{\includegraphics{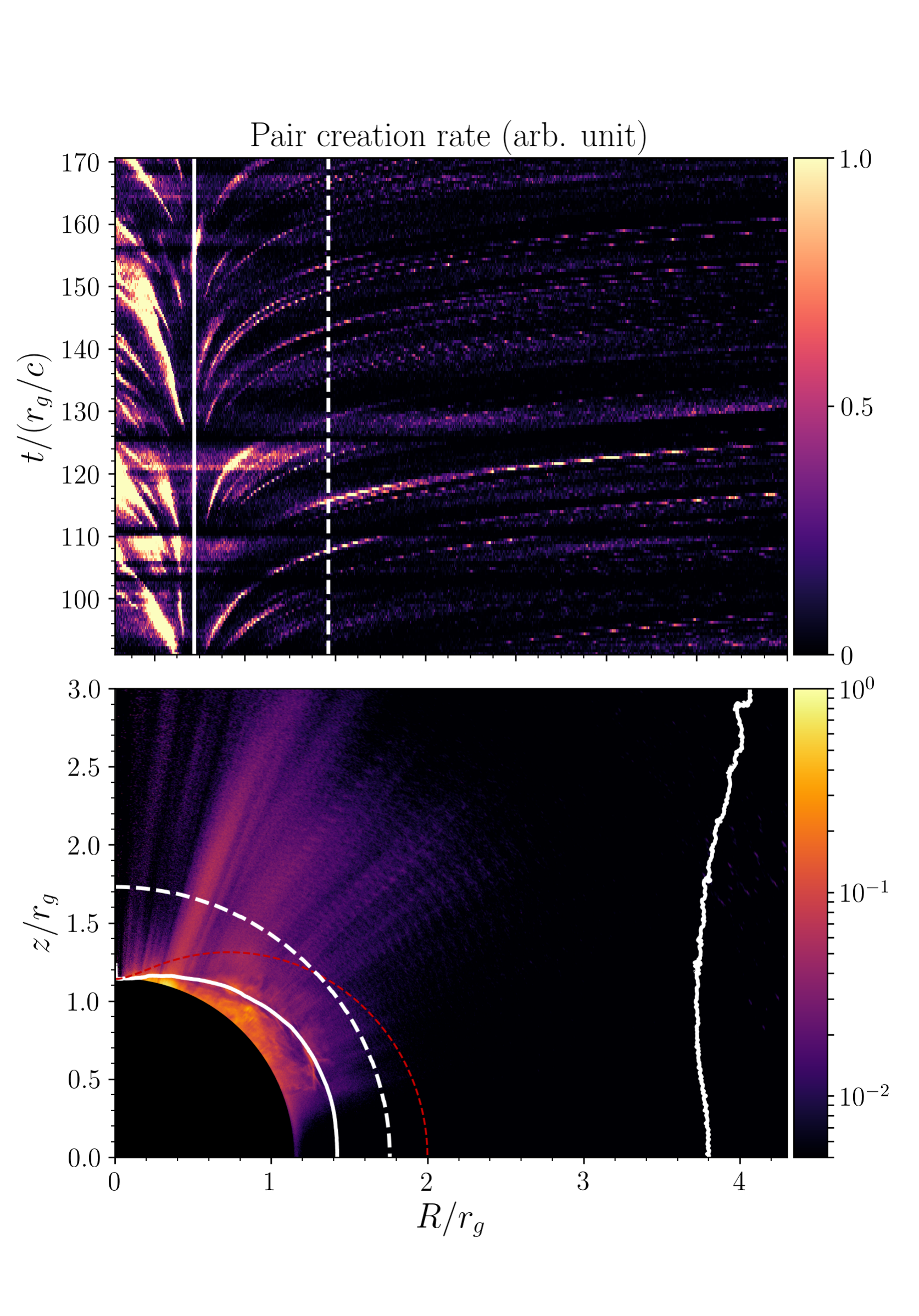}}
    \vspace{-1.0cm}
	\caption{Top panel: spacetime diagram of the pair creation rate at $\theta=\pi/4$ for the high opacity simulation, in arbitrary units. The white solid (resp. dashed) line marks the location of the inner light surface (resp. the null surface) at $\theta=\pi/4$. Although pair creation is continuous in time in the simulations, trajectories look discretized because of downsampling. Bottom panel: 2D map of the time-averaged pair creation rate. The white solid (resp. dashed) line marks the time-averaged reconstructed light surfaces (resp. analytical null surface).}
	\label{fig:pair_creation}
\end{figure}

Our simulations have $\vec{\Omega} \cdot \vec{B} > 0$ in the upper hemisphere and $\vec{\Omega} \cdot \vec{B} < 0$ in the lower one, where $\vec{\Omega}$ is the black-hole angular velocity vector. In order to screen the electric field, the black-hole magnetosphere requires a negative poloidal current in the upper hemisphere $(z>0)$ and a positive current in the lower hemisphere $(z<0)$. Electron density is always greater than the positron density for $z>0$, and lower for $z<0$. Still, the plasma remains globally neutral during the simulation. A species in the upper hemisphere has the same behavior as its anti-species in the lower hemisphere. Particles flow mainly radially, along the magnetic field lines. We ran four simulations with $\tau_0=5$, $10$, $20$ and $30$. A steady state is reached after around $50$ to $100$ $r_g /c$, as determined by the total number of particles in the box.

We observe a transition between two regimes with increasing $\tau_0$ (Fig.~\ref{fig:densities}). At low optical depths ($\tau_0\lesssim 10$), pair formation occurs far from the black hole, resulting in a macroscopic low-density zone close to the horizon (left panel of Fig.~\ref{fig:densities}). The electric field remains unscreened in this zone, so a large and steady gap forms. Particles experience the full vacuum potential which puts them deep into the Klein-Nishina regime. This results in a drop in the IC cross section, pushing IC emission, and hence pair production, even further outwards, outside of the acceleration zone. In this regime, acceleration and pair creation are spatially decoupled. At even lower opacity the gap is so large and the particle energy so high that all particles escape the simulation before pair producing.



At high optical depths ($\tau_0 \gtrsim 30$), on the other hand, the gap is narrow. Pairs are created at low altitudes so the gap can be screened efficiently. It is extremely intermittent, ejecting shreds of pair plasma outwards (see the inset of Fig~\ref{fig:densities} for $\tau_0=30$). After a burst of pair creation, a significant number of positrons are expelled, with the help of positive wiggles of the unscreened electric field (see Fig.~\ref{fig:3}). The typical normalized value of the unscreened electric field as measured by FIDOs, $\vec{D} \cdot \vec{B} / B^2$, ranges between $10^{-3}$ and $10^{-2}$, which is similar to the \textit{ad hoc} values used in \citetalias{Parfrey_2019}. Intermediate opacity simulations display an intermediate regime: high latitude field lines behave similarly to the low opacity case (see Fig.~\ref{fig:densities} for $\tau_0=20$), whereas field lines close to the equator show the same time-dependent behavior as the high opacity run. The inner and outer light surfaces, beyond which the rotation of magnetic field lines is superluminal~\citep{Komissarov_2004a}, are shown on the lower plots in Fig.~\ref{fig:densities}. Their shapes at high opacity are consistent with what was previously derived in the force-free regime \citep[e.g.][]{Komissarov_2004a,Nathanail_2014}. The size of the simulation box was set so as to include both light surfaces.

The insets in the lower panels of Fig. \ref{fig:densities} show the radial component of the electron $3$-velocity near the horizon. Focusing on the upper hemisphere only, in all simulations there is an electron velocity separation surface located exactly at the inner light surface. The positron velocity separation surface has a different location, which depends on the opacity. It always lies between the inner and outer light surfaces. The higher $\tau_0$, the closer to the black hole the positron separation surface is. The situation is symmetric (switching positrons and electrons) in the lower hemisphere. The high opacity simulations present similarities with the low plasma supply simulation in \citetalias{Parfrey_2019}, in particular regarding the role of the light surface. However, in our simulations all particles fly away from the black hole outside of the outer light surface, as a result of the different magnetic configuration used. Within the inner light surface, both species fall into the black hole for all $\tau_0$. We ran a simulation at high opacity but with spin $a=0.75$ and confirmed that the inner light surface retains the same role.

The MHD stagnation surface, separating inflow and outflow in single-fluid MHD~\citep{Takahashi_1990}, has been suggested as a plausible position for the gap~\citep{Broderick_2015}. Its location can be derived analytically~\citep{Hirotani2016} and is presented in the top panel of Fig.~\ref{fig:densities}. The null surface is where the general-relativistic Goldreich-Julian charge density vanishes \citep{Levinson_2017} and has also been proposed as as plausible gap position. We find that both the stagnation surface and the null surface are irrelevant for the pair discharge, and that the inner light surface is where the gap forms. As the gap opens, a burst of unscreened electric field either plunges inside the hole or moves outwards. Subsequent pair creation occurs in this burst as it propagates, populating the magnetosphere with pair plasma. This is visible in the upper panel in Fig.~\ref{fig:pair_creation}, which shows a spacetime diagram of the pair creation rate. This highlights the variability of the gap as well as its small spatial extent.

A typical sequence of bursts from the high opacity simulations is shown in Fig.~\ref{fig:3}. The electrostatic gap that opens accelerates particles, which produce high-energy photons that soon pair produce high-energy particles. As these secondary particles are created, they gradually screen the electric field parallel to the field lines. The intensity and duration of the bursts are highly variable. They have a spatial extent of a fraction of $r_g$ (see Fig. \ref{fig:3}), which appears promising for interpreting ultra-fast variability of AGN. We find that the gap size is controlled by the IC mean free path. At high opacity, the gap width is comparable to the IC mean free path in the Thomson regime $r_g/\tau_0$. The gap width, measured with the unscreened electric field, is $\sim 0.06 r_g$ at $\tau_0=30$. At low opacity, the mean free path becomes comparable to $r_g$. Particles reach high Lorentz factors in the gap, so the IC cross section drops, further increasing the mean free path. 

The multiplicity of the plasma flow is high in the gap (around $10$), and reaches $2$ outside of a burst. The high opacity solution is already very close to being force-free. We observed that the whole magnetosphere, despite being time-dependent due to the bursts, rotates consistently at the optimal predicted angular velocity $\approx \omega_\mathrm{BH}/2$ for a force-free magnetosphere~\citep{Blandford_1977, Komissarov_2004a}, except at low optical depth where we observe significant deviations. Going to higher $\tilde{B}_0 \tilde{\varepsilon}_0$ would likely increase the multiplicity and allow the magnetosphere to be even more force-free. The total Poynting power output measured in the simulations is also consistent with the BZ prediction~\citep{Blandford_1977, Tchekhovskoy_2010} $L_{\mathrm{BZ}}=B_0^2 \omega_\mathrm{BH}^2/6$ at all opacities (see the figures in the Supplemental Material). This supports the role of the BZ mechanism in the extraction of energy from the black hole, and the possibility that IC scattering and $\gamma\gamma$ pair production processes can supply sufficient plasma to activate this mechanism. 

At low opacity a sizeable fraction of the Poynting flux (around $20\%$) is dissipated within the numerical box. A large fraction of the dissipated energy goes into high-energy photons and leptons. The bulk energy-at-infinity of the leptons within the ergosphere can be negative, as emphasized in \citetalias{Parfrey_2019}; we find that they significantly contribute to energy extraction from the black hole at low opacity. At higher opacity dissipation is smaller since the gap is narrow. The energy flux carried by leptons becomes negligible\footnote{This does not contradict the conclusion, obtained in \citetalias{Parfrey_2019}, that particles with negative energy-at-infinity can contribute significantly to black-hole energy extraction. In their study, most of them were located in a current sheet, while there is none in our simulations.}. The dissipated energy is rather deposited in photons below the pair creation threshold, which we remove from the simulation to save computing time. The power carried by these photons can be estimated by computing the dissipation rate $\int_\mathcal{V} E_i J^i \diff{V}$ integrated over the whole simulation box. At high optical depths, the dissipated power is around $3\%$ of the output Poynting flux. Therefore these bursts are likely to come with gamma-ray emission, possibly detectable from Earth. 




\begin{figure}[t!]
    \centering
    \resizebox{\hsize}{!}{\includegraphics{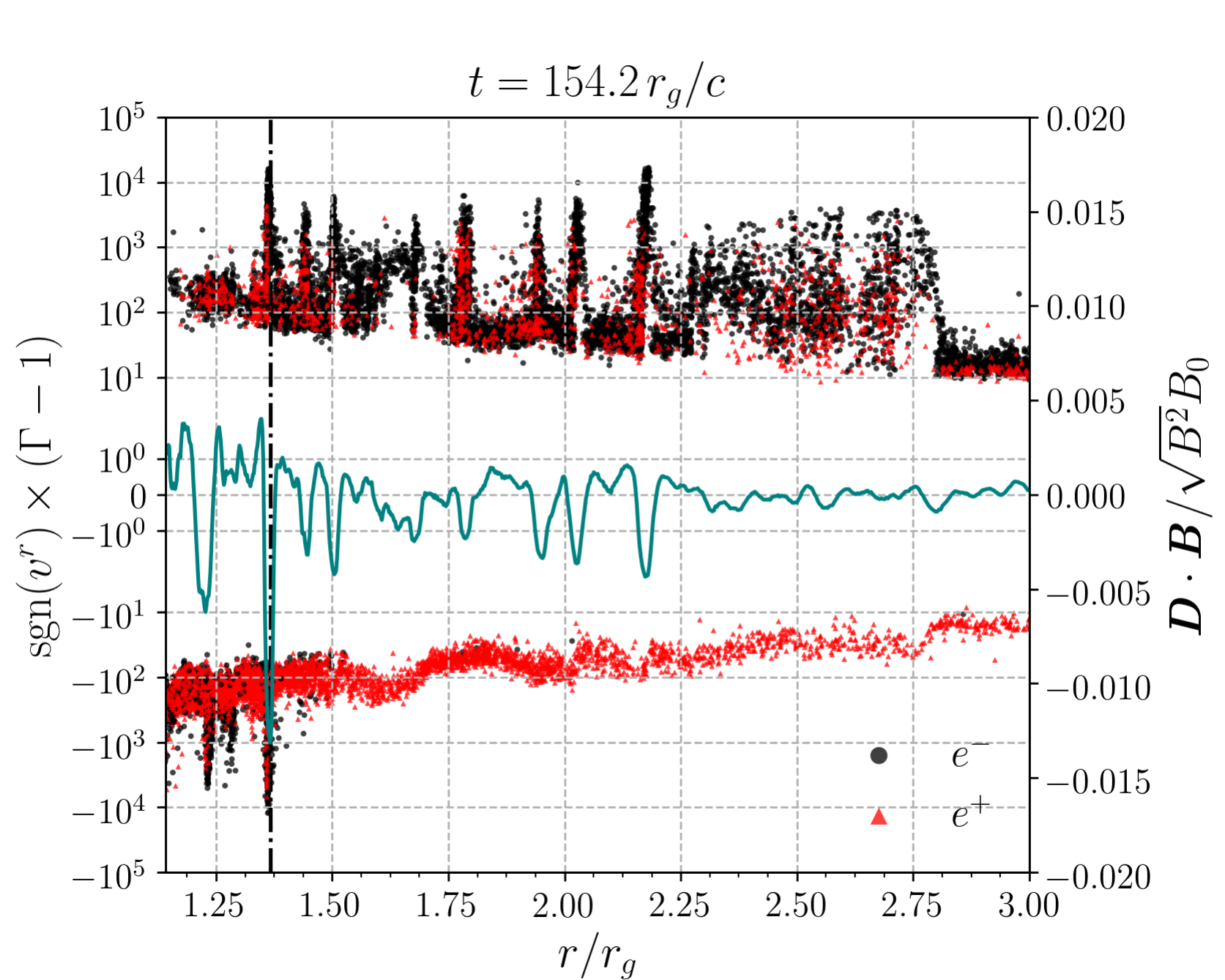}}
	\caption{Snapshot of the phase space for electrons (black dots) and positrons (red triangles) sampled at $\theta_0=\pi/4 \pm 0.02$ during a burst, for $\tau_0=30$. Particles are denoted by $\mathrm{sgn}(v^r) ( \Gamma -1)$, where $\Gamma$ is the FIDO-measured Lorentz factor and $v^r$ is their radial $3$-velocity. The blue solid line is the normalized unscreened electric field profile at $\theta_0$. The vertical dash-dotted line marks the location of the light surface at $\theta_0$. For clarity, only $20\%$ of the particles are displayed.}
	\label{fig:3}
\end{figure}


Our results show some similarities with 1D models, but also important differences which justify the need for multi-dimensional simulations. Similarly to \citet{Chen_2019}, we find that the gap opens quasi-periodically. However, unlike them we find that discharges happen at the inner light surface, whereas the null surface seems to play no role. Additionally, while their gap has a size $\gtrsim r_g$, we find that the gap size is much smaller than the black hole size in the high optical depth regime (although it remains much larger than the plasma skin depth). A major difference between 1D and 2D is that field lines do not all behave as a coherent entity. Therefore the pair creation bursts have a smaller spatial extent and the time variability is higher in our simulations than in 1D models. On the other hand, we do not observe the quasi-steady, noisy state obtained by \citet{Levinson_2018}, or by \citet{Chen_2019} at low resolution. This might be because field lines can still weakly interact through the electric field in the $(\theta,\varphi)$ plane, retaining some coherence at small scale.

In a future work we will aim to reproduce radio and gamma-ray observations of AGN, by applying the self-consistent radiative transfer treatment used in this study to other magnetic configurations. Although the structure of the outflow might be quantitatively different, the inner light surface is not expected to depend significantly on the large-scale magnetic configuration and therefore the broad conclusions we draw from this study should hold generally.



\vspace{0.5cm}

\begin{acknowledgements}

The authors would like to thank A. Levinson, M. Medvedev, V. Beskin and E. Quataert for useful discussions. This work has been supported by the Programme National des Hautes Énergies of CNRS/INSU, CNES, the France-Berkeley Fund (Project \#24-2018). Computing resources were provided by TGCC and CINES under the allocation A0050407669 made by GENCI, and by Scientific Computing Core at Flatiron Institute. B. Crinquand wants to acknowledge the 2019 Summer School at the Center for Computational Astrophysics, Flatiron Institute, where part of this research has been performed. Research at the Flatiron Institute is supported by the Simons Foundation, which also supported KP. We also thank the anonymous referees for valuable comments on the manuscript.

\end{acknowledgements}



\widetext
\clearpage

\begin{center}
\textbf{\large Multi-dimensional simulations of ergospheric pair discharges around black holes:\\ 
Supplemental Material}
\end{center}

\section{Magnetic configuration}

The initial electromagnetic field in our simulations is prescribed by the following $4$-potential~\citep{Novikov}, written in Kerr-Schild spherical coordinates $(t,r,\theta,\varphi)$:

\begin{equation} \label{eq:a_mu}
    A_\mu= B_0 r_g \left( \dfrac{a \cos{\theta}}{(r/r_g)^2 + a^2 \cos^2{\theta}}, 0 , 0 , - \dfrac{(r/r_g)^2 + a^2}{(r/r_g)^2 + a^2 \cos^2{\theta}} \cos{\theta} \right),
\end{equation}
where $a\in [0,1[$ is the dimensionless spin parameter of the black hole, $r_g$ the gravitational radius and $B_0$ the strength of the magnetic field. Eq.~\eqref{eq:a_mu} describes a solution to Maxwell's equations for a black hole with a magnetic monopole. The electromagnetic fields are then derived from $A_\mu$:
\begin{eqnarray}
B^r & = & \dfrac{1}{\sqrt{h}}\partial_\theta A_\varphi, \\
B^\theta & = & - \dfrac{1}{\sqrt{h}}\partial_r A_\varphi, \\
B^\varphi & = & 0, \\
E_r & = & \partial_r A_t, \\
E_\theta & = & \partial_\theta A_t, \\
E_\varphi & = & \partial_\varphi A_t = 0, 
\end{eqnarray}
where $h$ is the determinant of the spatial $3$-metric. We verified that the vacuum electromagnetic field relaxes to the solution described by Eq.~\eqref{eq:a_mu} if we start with a purely radial magnetic field and no electric field.

\section{Radiative transfer}

We include two radiative processes in our model: inverse Compton (IC) scattering and photon-photon pair production. We introduce ``fiducial observers'' (FIDOs), whose wordlines are orthogonal to spatial hypersurfaces. We take advantage of the fact that FIDOs are locally inertial observers, so that the laws of special relativity can be applied, provided we only use FIDO-measured physical quantities. For simplicity, we assume the soft background radiation field to be isotropic, mono-energetic, and uniform, with density $n_0$ in the FIDOs' rest frame. We also neglect pairs that would be produced by the annihilation of the soft background radiation field on itself, i.e. due to MeV emission from the radiatively-inefficient flow. The density of pairs created through this process is usually expected to be much smaller than the Goldreich-Julian density, and therefore too low to screen the gap, for the very low accretion rate found for M87*~\citep{Levinson_2011,Moscibrodzka_2011}.

\subsection{Condition for interaction}

The opacity of IC scattering, for a lepton of Lorentz factor $\gamma=(1-\beta^2)^{-1/2}$ propagating in the soft radiation field, is computed as~\citep{Blumenthal_1970}
\begin{equation}
 \kappa_{IC}(\gamma) = \dfrac{\tau_0}{2 r_g} \int_{-\pi}^{\pi} \diff{\theta} \sin{\theta} (1- \beta \cos{\theta} ) \sigma_\mathrm{KN} (\varepsilon_0, \gamma, \theta),
\end{equation}
where $\sigma_\mathrm{KN}$ is the Klein-Nishina cross section, and $\tau_0 = n_0 r_g \sigma_T$ is the fiducial opacity ($\sigma_T$ is the Thomson cross section). The pair production opacity is computed similarly, using the pair production cross-section $\sigma_{\gamma\gamma}$ instead of $\sigma_\mathrm{KN}$. The optical depth traversed by a particle whose spatial coordinates have changed by an amount $\diff{x}^i$ is measured during a time step as
\begin{equation}
 \delta \tau = \kappa \sqrt{h_{ij} \diff{x}^i \diff{x}^j},
\end{equation}
where $h_{ij}$ is the spatial $3$-metric. A number $p$ is randomly drawn with uniform probability between $0$ and $1$; a scattering event occurs provided $p < 1- \exp(-\delta \tau)$. 


\subsection{Inverse Compton scattering}

We consider, in the FIDO frame, a lepton of energy $\gamma m_e c^2$ interacting with a soft photon of energy $\varepsilon_0$. The photon makes angles $(\theta_0,\varphi_0)$ with the lepton velocity. In the following, quantities defined in the lepton rest frame will be primed. After the scattering, the photon energy is $\varepsilon_1$. The energy of the photon in the lepton rest frame is $\varepsilon_0'=\varepsilon_0 \gamma (1-\beta \mu_0)$, where $\mu_0 = \cos{\theta_0}$ and $\beta=(1-\gamma^{-2})^{1/2}$. The kinematics of IC scattering yield 
\begin{equation} \label{eq:ICkinematics}
    \varepsilon_1' = \dfrac{\varepsilon_0'}{1+\dfrac{\varepsilon_0'}{m_e c^2} (1- \cos \Theta')},
\end{equation}
where $\cos{\Theta'}= \mu_0' \mu_1' + \sqrt{1-\mu_0'^2} \sqrt{1-\mu_1'^2} \cos{(\varphi_1' - \varphi_0')}$, $\Theta'$ being the angle between the incoming and the scattered photon directions in the lepton rest frame. We assume that the lepton is very energetic ($\gamma \gg 1$), so we have $\mu_0' \approx -1$ by virtue of relativistic beaming. We can therefore approximate $\cos{\Theta'} \approx - \mu_1'$. The energy of the scattered photon in the lepton rest frame $\varepsilon_1'$ is determined using the full IC differential cross-section from quantum electrodynamics (QED). Given $\varepsilon_1'$, the scattering angle in this frame is deduced using Eq.~\eqref{eq:ICkinematics}:
\begin{equation} \label{eq:angle_Lorentz}
    \mu_1'=\dfrac{m_e c^2}{\varepsilon_1'} - \dfrac{m_e c^2}{\varepsilon_0'} -1.
\end{equation}
Finally, another Lorentz transformation gives the energy of the scattered photon back in the FIDO frame: 
\begin{equation} \label{eq:energy_Lorentz}
    \varepsilon_1 = \gamma (1+ \beta \mu_1') \varepsilon_1'.
\end{equation}
Thus, once the angle of the incoming photon $\theta_0$ is randomly drawn, we only need to draw the energy of the scattered photon in the lepton rest frame $\varepsilon_1'$ from QED.

We can now summarize our Monte-Carlo scheme for IC scattering. First the FIDO-measured Lorentz factor $\gamma_0=\sqrt{1+ h^{jk} u_j u_k}$ of a lepton is computed. Since the radiation field is isotropic, we draw uniformly the random variable $\mu_0 \in [-1,1]$ and use it to compute $\varepsilon_0'$ by a Lorentz transformation. The scattered photon energy $\varepsilon_1'$ is then determined using the full IC differential cross-section. Then $\mu_1'$ is given by Eq.~\eqref{eq:angle_Lorentz}, and we deduce the energy of the scattered photon in the FIDO frame with Eq.~\eqref{eq:energy_Lorentz}. In the code, we create a high-energy photon at the location of the scattering lepton, with energy $\varepsilon_1$. Assuming strong relativistic beaming again, the direction of the scattered photon in the FIDO frame is the same as that of the scattering lepton. The new Lorentz factor of the lepton is $\gamma_1=\gamma_0 + \varepsilon_0 / m_e c^2 - \varepsilon_1 / m_e c^2$, using energy conservation.

\citet{Jones_1968} derived the analytical photon spectrum scattered by a single lepton bathed in a uniform, isotropic and mono-energetic radiation field, which is valid both in the Thomson and the Klein-Nishina regimes. We confirmed that the photon spectrum obtained in our numerical simulations matched this analytical prediction in both regimes.

\subsection{Pair production}

We consider two photons of energies $\varepsilon_0$ and $\varepsilon_1$ in the FIDO frame, colliding with an angle $\theta_0$. In the following, we will assume that $\varepsilon_0 \ll \varepsilon_1$, where $\varepsilon_0$ is the energy of a soft photon from the background radiation field. An electron/positron ($e^\pm$) pair can only be created provided~\citep{Gould_1967}
\begin{equation} \label{eq:s}
    s=\dfrac{1}{2} \varepsilon_0 \varepsilon_1 (1-\cos{\theta_0}) \ge (m_e c^2)^2.
\end{equation}
In the following, quantities defined in the center-of-mass (COM) frame of the pair will be primed. In the limit $\varepsilon_1 \gg \varepsilon_0$, the Lorentz factor and velocity of the COM frame with respect to the FIDO frame are $\gamma_{CM} \approx \varepsilon_1 / 2 \sqrt{s}$ and $\beta_{CM}=1- 2s / \varepsilon_1^2$~\citep{Bonometto_1971}. The electron and the positron both have the same energy $\gamma_1' m_e c^2 = \sqrt{s}$ in the COM frame. The angle $\theta_1'$, at which the produced pair propagates with respect to the gamma-ray direction in the COM frame, is determined by QED. Once $\mu_1'=\cos{\theta_1'}$ is known, the energy of the electron is given by a Lorentz transformation back to the FIDO frame:
\begin{equation} \label{eq:energy-}
    \gamma_- = \gamma_{CM}(\gamma_1'+ \beta_{CM} \mu_1' \sqrt{{\gamma_1'}^2 -1}),
\end{equation}
whereas the positron energy is determined by energy conservation:
\begin{equation} \label{eq:energy+}
    \gamma_+ m_e c^2 = \varepsilon_0 + \varepsilon_1 - \gamma_- m_e c^2 \approx \varepsilon_1 - \gamma_- m_e c^2.
\end{equation}
Since the energy distribution is symmetric with respect to $\varepsilon_1 /2$, we arbitrarily choose to pick the electron first. 

We can summarize our Monte-Carlo scheme for pair production. First the FIDO-measured energy of a gamma photon $\varepsilon_1=\sqrt{h^{jk} u_j u_k}$ is computed. We draw uniformly the random variable $\mu_0 = \cos \theta_0 \in [-1,1]$ and use it to compute $s$ from Eq.~\eqref{eq:s}. If $s \le (m_e c^2)^2$ then no pair is created. Otherwise we compute $\gamma_{CM}$, $\beta_{CM}$, $\gamma_1'$, and then draw $\mu_1'$ using the QED differential cross-section for pair creation. The gamma photon is discarded from our simulation, and an $e^\pm$ pair is created in its place, with the energies of the electron and the positron given respectively by Eq.~\eqref{eq:energy-} and~\eqref{eq:energy+}. We take the direction of propagation of the created pair to be along that of the primary gamma-ray. This approximation is valid provided $\gamma_{CM} \gg 1$, which always holds since $\varepsilon_1 \gg \varepsilon_0$. 

\citet{Aharonian_1983} derived the analytical pair spectrum for a high-energy photon propagating in an isotropic and mono-energetic radiation field, in the case where the high-energy photon has an energy much greater than that of a photon from the background field. We verified that the agreement between this analytical prediction and the output of the algorithm is good, both close to the pair creation threshold $(s\approx1)$, where the electron and the positron have the same energy in the FIDO frame, and far from the threshold ($\gamma_+ \approx \varepsilon_1$ or $\gamma_- \approx \varepsilon_1$), where the pair's energy is asymmetric.

\section{Poynting flux}

\begin{figure*}[b!]
    \centering
    \captionsetup[subfigure]{position=top,textfont=normalfont,singlelinecheck=off,justification=raggedright}
	\makebox[\textwidth][c]{
	\sidesubfloat[]{\includegraphics[width=8.0cm]{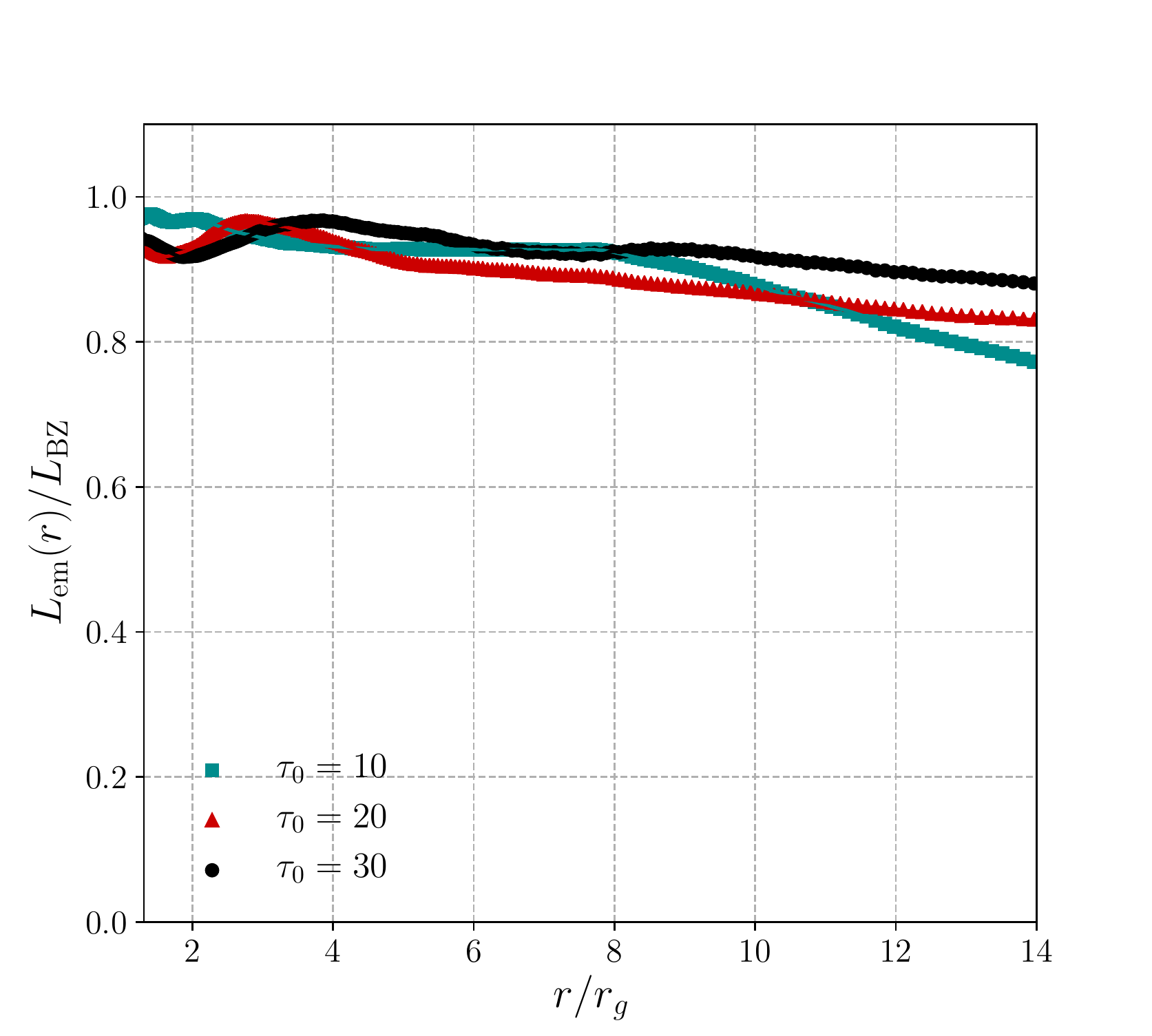}\label{fig:poynting}} 
	\sidesubfloat[]{\includegraphics[width=8.0cm]{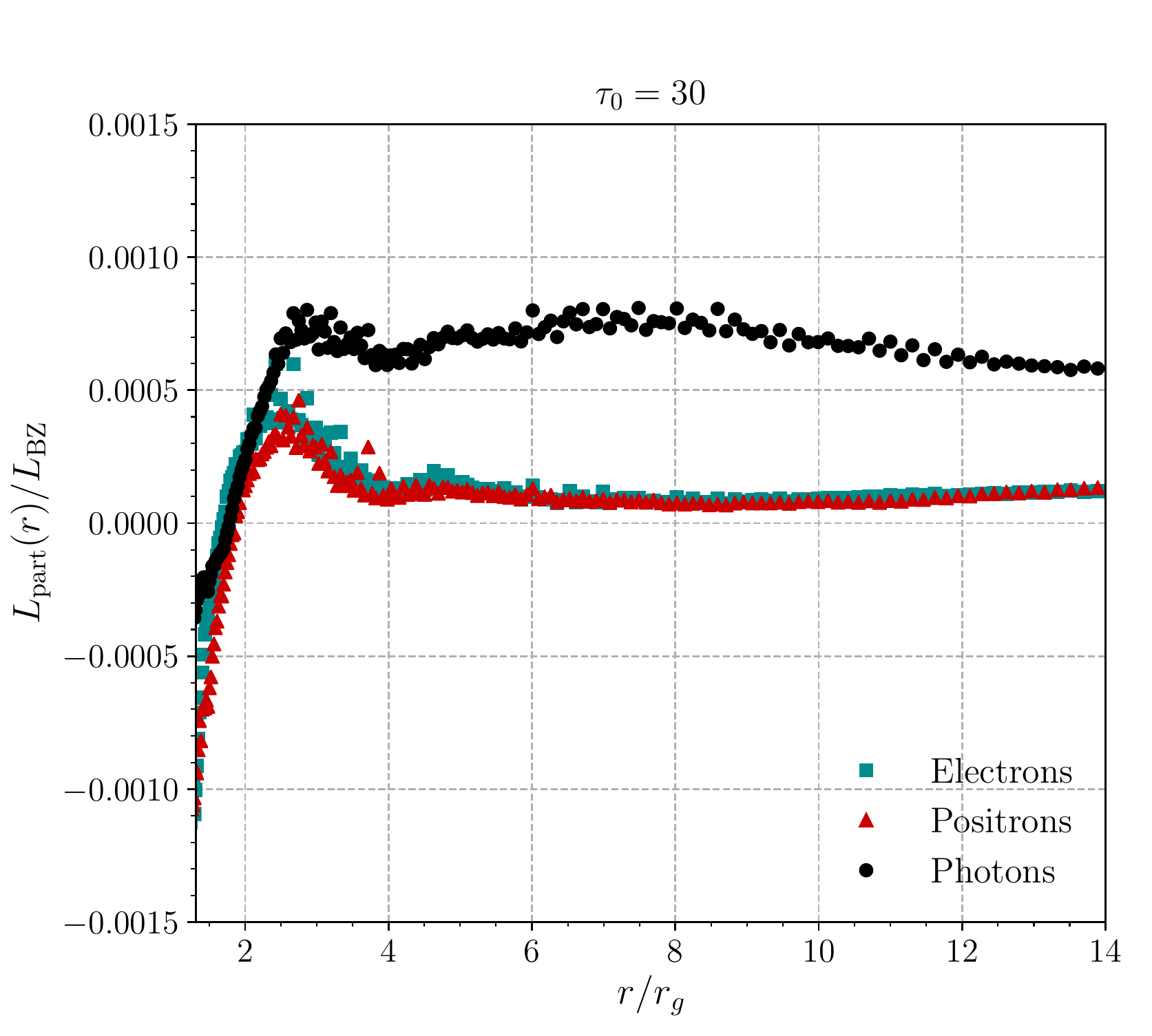}\label{fig:particles}}} 
	\caption{(a) Steady-state Poynting flux through spherical shells centered on the black hole for three optical depths, $\tau_0=10$, $20$ and $30$. (b) Electron, positron and photon energy-at-infinity flux through spherical shells centered on the black hole for $\tau_0=30$. All fluxes are scaled with $L_\mathrm{BZ}$.}
    \label{fig:fig}
\end{figure*}

Fig.~\ref{fig:poynting} shows the total Poynting flux through spheres centered on the black hole, as a function of the radius of that sphere. The fluxes are normalized with the total power output of the black hole predicted by the BZ mechanism~\citep{Blandford_1977}:
\begin{equation}
    L_{\mathrm{BZ}}=\dfrac{B_0^2 \omega_\mathrm{BH}^2}{6},
\end{equation}
where $\omega_\mathrm{BH}=(ca/2r_g)/(1+\sqrt{1-a^2})$ is the angular velocity of the black hole.
This expression is accurate to second order in $\omega_\mathrm{BH}$~\citep{Tchekhovskoy_2010}. The Poynting flux decreases with increasing $r$ because some energy is dissipated in the gap and converted into lepton kinetic energy. Dissipation of the Poynting flux is larger at lower opacity since the non-ideal gap region is wider. Fig.~\ref{fig:particles} shows the energy-at-infinity fluxes carried by electrons, positrons and high-energy photons (above the pair creation threshold) in the high-opacity simulation. Their contribution to black-hole energy extraction is very small. At high opacity, the dissipated electromagnetic energy is mostly transferred to photons below the pair creation threshold. 

\section{Angular velocity of the field lines}

The field lines' angular velocity can be evaluated as~\citep{Blandford_1977,Komissarov_2007} $\Omega_\mathrm{F}=- E_\theta / \sqrt{h} B^r$. Fig.~\ref{fig:3} shows that the whole magnetosphere rotates consistently at $\Omega_\mathrm{F} \approx \omega_\mathrm{BH}/2$~\citep{Blandford_1977}, 
except at very low opacity ($\tau_0=5$ in Fig.~\ref{fig:3}), when the magnetosphere is not densely filled with pair plasma and is far from the force-free solution.

\begin{figure}[h!]
    \centering
    \includegraphics[width=9.0cm]{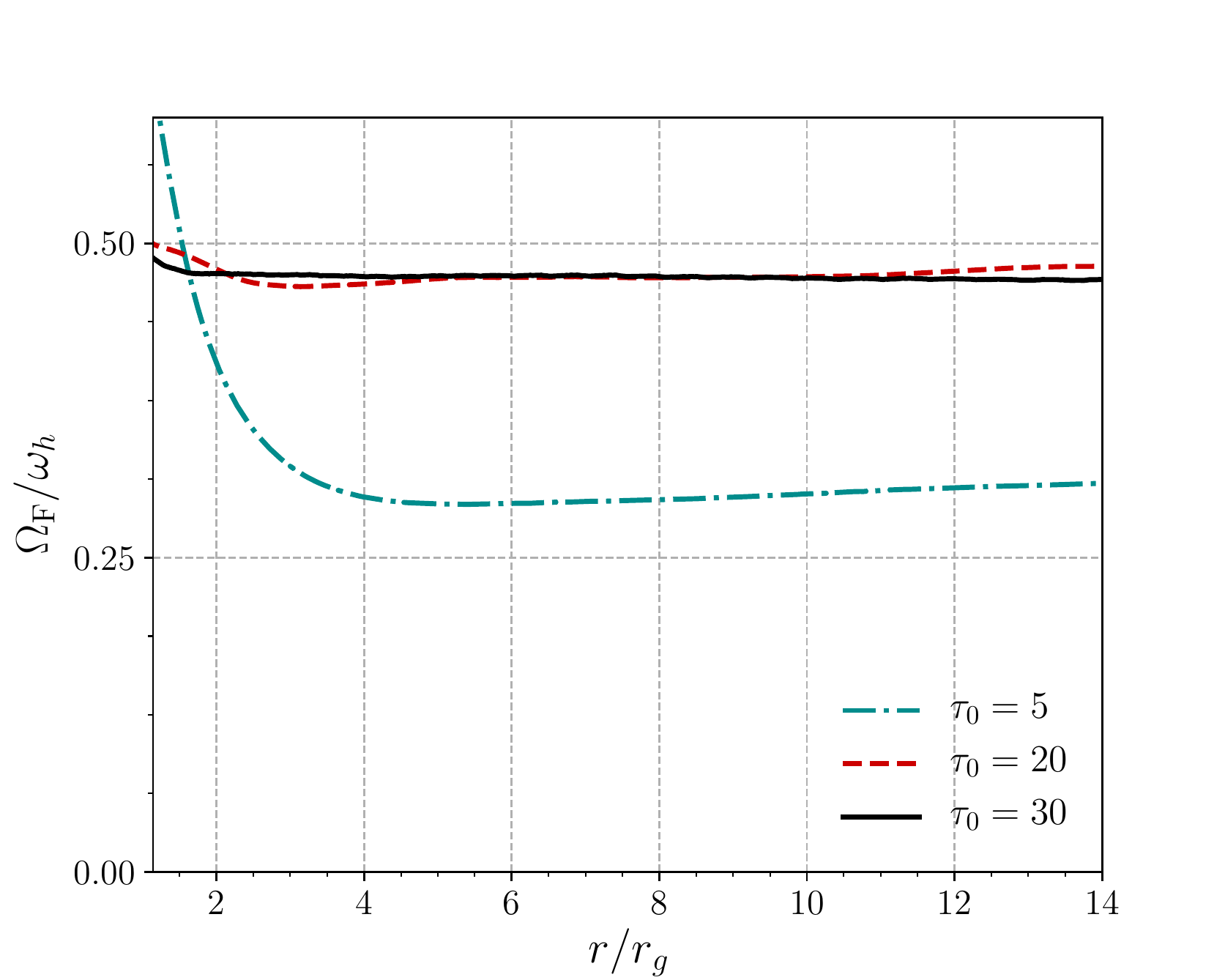}
	\caption{Angular velocity of the field lines, averaged over $\theta$, as a function of $r$, for three optical depths, $\tau_0=5$, $20$ and $30$.}
	\label{fig:3}
\end{figure}

\bibliographystyle{apsrev4-1}

\bibliography{biblio}

\begin{thebibliography}{33}%
\makeatletter
\providecommand \@ifxundefined [1]{%
 \@ifx{#1\undefined}
}%
\providecommand \@ifnum [1]{%
 \ifnum #1\expandafter \@firstoftwo
 \else \expandafter \@secondoftwo
 \fi
}%
\providecommand \@ifx [1]{%
 \ifx #1\expandafter \@firstoftwo
 \else \expandafter \@secondoftwo
 \fi
}%
\providecommand \natexlab [1]{#1}%
\providecommand \enquote  [1]{``#1''}%
\providecommand \bibnamefont  [1]{#1}%
\providecommand \bibfnamefont [1]{#1}%
\providecommand \citenamefont [1]{#1}%
\providecommand \href@noop [0]{\@secondoftwo}%
\providecommand \href [0]{\begingroup \@sanitize@url \@href}%
\providecommand \@href[1]{\@@startlink{#1}\@@href}%
\providecommand \@@href[1]{\endgroup#1\@@endlink}%
\providecommand \@sanitize@url [0]{\catcode `\\12\catcode `\$12\catcode
  `\&12\catcode `\#12\catcode `\^12\catcode `\_12\catcode `\%12\relax}%
\providecommand \@@startlink[1]{}%
\providecommand \@@endlink[0]{}%
\providecommand \url  [0]{\begingroup\@sanitize@url \@url }%
\providecommand \@url [1]{\endgroup\@href {#1}{\urlprefix }}%
\providecommand \urlprefix  [0]{URL }%
\providecommand \Eprint [0]{\href }%
\providecommand \doibase [0]{http://dx.doi.org/}%
\providecommand \selectlanguage [0]{\@gobble}%
\providecommand \bibinfo  [0]{\@secondoftwo}%
\providecommand \bibfield  [0]{\@secondoftwo}%
\providecommand \translation [1]{[#1]}%
\providecommand \BibitemOpen [0]{}%
\providecommand \bibitemStop [0]{}%
\providecommand \bibitemNoStop [0]{.\EOS\space}%
\providecommand \EOS [0]{\spacefactor3000\relax}%
\providecommand \BibitemShut  [1]{\csname bibitem#1\endcsname}%
\let\auto@bib@innerbib\@empty
\bibitem [{\citenamefont {{Walker}}\ \emph {et~al.}(2018)\citenamefont
  {{Walker}}, \citenamefont {{Hardee}}, \citenamefont {{Davies}}, \citenamefont
  {{Ly}},\ and\ \citenamefont {{Junor}}}]{Walker_2018}%
  \BibitemOpen
  \bibfield  {author} {\bibinfo {author} {\bibfnamefont {R.~C.}\ \bibnamefont
  {{Walker}}}, \bibinfo {author} {\bibfnamefont {P.~E.}\ \bibnamefont
  {{Hardee}}}, \bibinfo {author} {\bibfnamefont {F.~B.}\ \bibnamefont
  {{Davies}}}, \bibinfo {author} {\bibfnamefont {C.}~\bibnamefont {{Ly}}}, \
  and\ \bibinfo {author} {\bibfnamefont {W.}~\bibnamefont {{Junor}}},\ }\href
  {\doibase 10.3847/1538-4357/aaafcc} {\bibfield  {journal} {\bibinfo
  {journal} {\apj}\ }\textbf {\bibinfo {volume} {855}},\ \bibinfo {eid} {128}
  (\bibinfo {year} {2018})},\ \Eprint {http://arxiv.org/abs/1802.06166}
  {arXiv:1802.06166 [astro-ph.HE]} \BibitemShut {NoStop}%
\bibitem [{\citenamefont {{Abramowski}}\ \emph {et~al.}(2012)\citenamefont
  {{Abramowski}}, \citenamefont {{Acero}}, \citenamefont {{Aharonian}},
  \citenamefont {{Akhperjanian}}, \citenamefont {{Anton}}, \citenamefont
  {{Balzer}}, \citenamefont {{Barnacka}}, \citenamefont {{Barres de Almeida}},
  \citenamefont {{Becherini}}, \citenamefont {{Becker}} \emph
  {et~al.}}]{Abramowski_2012}%
  \BibitemOpen
  \bibfield  {author} {\bibinfo {author} {\bibfnamefont {A.}~\bibnamefont
  {{Abramowski}}}, \bibinfo {author} {\bibfnamefont {F.}~\bibnamefont
  {{Acero}}}, \bibinfo {author} {\bibfnamefont {F.}~\bibnamefont
  {{Aharonian}}}, \bibinfo {author} {\bibfnamefont {A.~G.}\ \bibnamefont
  {{Akhperjanian}}}, \bibinfo {author} {\bibfnamefont {G.}~\bibnamefont
  {{Anton}}}, \bibinfo {author} {\bibfnamefont {A.}~\bibnamefont {{Balzer}}},
  \bibinfo {author} {\bibfnamefont {A.}~\bibnamefont {{Barnacka}}}, \bibinfo
  {author} {\bibfnamefont {U.}~\bibnamefont {{Barres de Almeida}}}, \bibinfo
  {author} {\bibfnamefont {Y.}~\bibnamefont {{Becherini}}}, \bibinfo {author}
  {\bibfnamefont {J.}~\bibnamefont {{Becker}}},  \emph {et~al.},\ }\href
  {\doibase 10.1088/0004-637X/746/2/151} {\bibfield  {journal} {\bibinfo
  {journal} {\apj}\ }\textbf {\bibinfo {volume} {746}},\ \bibinfo {eid} {151}
  (\bibinfo {year} {2012})},\ \Eprint {http://arxiv.org/abs/1111.5341}
  {arXiv:1111.5341 [astro-ph.CO]} \BibitemShut {NoStop}%
\bibitem [{\citenamefont {{Aharonian}}\ \emph {et~al.}(2006)\citenamefont
  {{Aharonian}}, \citenamefont {{Akhperjanian}}, \citenamefont {{Bazer-Bachi}},
  \citenamefont {{Beilicke}}, \citenamefont {{Benbow}}, \citenamefont
  {{Berge}}, \citenamefont {{Bernl{\"o}hr}}, \citenamefont {{Boisson}},
  \citenamefont {{Bolz}}, \citenamefont {{Borrel}} \emph
  {et~al.}}]{Aharonian_2006}%
  \BibitemOpen
  \bibfield  {author} {\bibinfo {author} {\bibfnamefont {F.}~\bibnamefont
  {{Aharonian}}}, \bibinfo {author} {\bibfnamefont {A.~G.}\ \bibnamefont
  {{Akhperjanian}}}, \bibinfo {author} {\bibfnamefont {A.~R.}\ \bibnamefont
  {{Bazer-Bachi}}}, \bibinfo {author} {\bibfnamefont {M.}~\bibnamefont
  {{Beilicke}}}, \bibinfo {author} {\bibfnamefont {W.}~\bibnamefont
  {{Benbow}}}, \bibinfo {author} {\bibfnamefont {D.}~\bibnamefont {{Berge}}},
  \bibinfo {author} {\bibfnamefont {K.}~\bibnamefont {{Bernl{\"o}hr}}},
  \bibinfo {author} {\bibfnamefont {C.}~\bibnamefont {{Boisson}}}, \bibinfo
  {author} {\bibfnamefont {O.}~\bibnamefont {{Bolz}}}, \bibinfo {author}
  {\bibfnamefont {V.}~\bibnamefont {{Borrel}}},  \emph {et~al.},\ }\href
  {\doibase 10.1126/science.1134408} {\bibfield  {journal} {\bibinfo  {journal}
  {Science}\ }\textbf {\bibinfo {volume} {314}},\ \bibinfo {pages} {1424}
  (\bibinfo {year} {2006})},\ \Eprint {http://arxiv.org/abs/astro-ph/0612016}
  {astro-ph/0612016} \BibitemShut {NoStop}%
\bibitem [{\citenamefont {{Event Horizon Telescope Collaboration}}\ \emph
  {et~al.}(2019{\natexlab{a}})\citenamefont {{Event Horizon Telescope
  Collaboration}}, \citenamefont {{Akiyama}}, \citenamefont {{Alberdi}},
  \citenamefont {{Alef}}, \citenamefont {{Asada}}, \citenamefont {{Azulay}},
  \citenamefont {{Baczko}}, \citenamefont {{Ball}}, \citenamefont
  {{Balokovi{\'c}}}, \citenamefont {{Barrett}} \emph {et~al.}}]{EHT_1}%
  \BibitemOpen
  \bibfield  {author} {\bibinfo {author} {\bibnamefont {{Event Horizon
  Telescope Collaboration}}}, \bibinfo {author} {\bibfnamefont
  {K.}~\bibnamefont {{Akiyama}}}, \bibinfo {author} {\bibfnamefont
  {A.}~\bibnamefont {{Alberdi}}}, \bibinfo {author} {\bibfnamefont
  {W.}~\bibnamefont {{Alef}}}, \bibinfo {author} {\bibfnamefont
  {K.}~\bibnamefont {{Asada}}}, \bibinfo {author} {\bibfnamefont
  {R.}~\bibnamefont {{Azulay}}}, \bibinfo {author} {\bibfnamefont {A.-K.}\
  \bibnamefont {{Baczko}}}, \bibinfo {author} {\bibfnamefont {D.}~\bibnamefont
  {{Ball}}}, \bibinfo {author} {\bibfnamefont {M.}~\bibnamefont
  {{Balokovi{\'c}}}}, \bibinfo {author} {\bibfnamefont {J.}~\bibnamefont
  {{Barrett}}},  \emph {et~al.},\ }\href {\doibase 10.3847/2041-8213/ab0ec7}
  {\bibfield  {journal} {\bibinfo  {journal} {\apj}\ }\textbf {\bibinfo
  {volume} {875}},\ \bibinfo {eid} {L1} (\bibinfo {year}
  {2019}{\natexlab{a}})},\ \Eprint {http://arxiv.org/abs/1906.11238}
  {arXiv:1906.11238} \BibitemShut {NoStop}%
\bibitem [{\citenamefont {{Blandford}}\ and\ \citenamefont
  {{Znajek}}(1977)}]{Blandford_1977}%
  \BibitemOpen
  \bibfield  {author} {\bibinfo {author} {\bibfnamefont {R.~D.}\ \bibnamefont
  {{Blandford}}}\ and\ \bibinfo {author} {\bibfnamefont {R.~L.}\ \bibnamefont
  {{Znajek}}},\ }\href {\doibase 10.1093/mnras/179.3.433} {\bibfield  {journal}
  {\bibinfo  {journal} {mnras}\ }\textbf {\bibinfo {volume} {179}},\ \bibinfo
  {pages} {433} (\bibinfo {year} {1977})}\BibitemShut {NoStop}%
\bibitem [{\citenamefont {{Levinson}}\ and\ \citenamefont
  {{Rieger}}(2011)}]{Levinson_2011}%
  \BibitemOpen
  \bibfield  {author} {\bibinfo {author} {\bibfnamefont {A.}~\bibnamefont
  {{Levinson}}}\ and\ \bibinfo {author} {\bibfnamefont {F.}~\bibnamefont
  {{Rieger}}},\ }\href {\doibase 10.1088/0004-637X/730/2/123} {\bibfield
  {journal} {\bibinfo  {journal} {\apj}\ }\textbf {\bibinfo {volume} {730}},\
  \bibinfo {eid} {123} (\bibinfo {year} {2011})},\ \Eprint
  {http://arxiv.org/abs/1011.5319} {arXiv:1011.5319 [astro-ph.HE]} \BibitemShut
  {NoStop}%
\bibitem [{\citenamefont {{Broderick}}\ and\ \citenamefont
  {{Tchekhovskoy}}(2015)}]{Broderick_2015}%
  \BibitemOpen
  \bibfield  {author} {\bibinfo {author} {\bibfnamefont {A.~E.}\ \bibnamefont
  {{Broderick}}}\ and\ \bibinfo {author} {\bibfnamefont {A.}~\bibnamefont
  {{Tchekhovskoy}}},\ }\href {\doibase 10.1088/0004-637X/809/1/97} {\bibfield
  {journal} {\bibinfo  {journal} {\apj}\ }\textbf {\bibinfo {volume} {809}},\
  \bibinfo {eid} {97} (\bibinfo {year} {2015})},\ \Eprint
  {http://arxiv.org/abs/1506.04754} {arXiv:1506.04754 [astro-ph.HE]}
  \BibitemShut {NoStop}%
\bibitem [{\citenamefont {{Hirotani}}\ and\ \citenamefont
  {{Pu}}(2016)}]{Hirotani2016}%
  \BibitemOpen
  \bibfield  {author} {\bibinfo {author} {\bibfnamefont {K.}~\bibnamefont
  {{Hirotani}}}\ and\ \bibinfo {author} {\bibfnamefont {H.-Y.}\ \bibnamefont
  {{Pu}}},\ }\href {\doibase 10.3847/0004-637X/818/1/50} {\bibfield  {journal}
  {\bibinfo  {journal} {\apj}\ }\textbf {\bibinfo {volume} {818}},\ \bibinfo
  {eid} {50} (\bibinfo {year} {2016})},\ \Eprint
  {http://arxiv.org/abs/1512.05026} {arXiv:1512.05026 [astro-ph.HE]}
  \BibitemShut {NoStop}%
\bibitem [{\citenamefont {{Levinson}}\ and\ \citenamefont
  {{Segev}}(2017)}]{Levinson_2017}%
  \BibitemOpen
  \bibfield  {author} {\bibinfo {author} {\bibfnamefont {A.}~\bibnamefont
  {{Levinson}}}\ and\ \bibinfo {author} {\bibfnamefont {N.}~\bibnamefont
  {{Segev}}},\ }\href {\doibase 10.1103/PhysRevD.96.123006} {\bibfield
  {journal} {\bibinfo  {journal} {\prd}\ }\textbf {\bibinfo {volume} {96}},\
  \bibinfo {eid} {123006} (\bibinfo {year} {2017})},\ \Eprint
  {http://arxiv.org/abs/1709.09397} {arXiv:1709.09397 [astro-ph.HE]}
  \BibitemShut {NoStop}%
\bibitem [{\citenamefont
  {{Komissarov}}(2004{\natexlab{a}})}]{Komissarov_2004b}%
  \BibitemOpen
  \bibfield  {author} {\bibinfo {author} {\bibfnamefont {S.~S.}\ \bibnamefont
  {{Komissarov}}},\ }\href {\doibase 10.1111/j.1365-2966.2004.07738.x}
  {\bibfield  {journal} {\bibinfo  {journal} {mnras}\ }\textbf {\bibinfo
  {volume} {350}},\ \bibinfo {pages} {1431} (\bibinfo {year}
  {2004}{\natexlab{a}})},\ \Eprint {http://arxiv.org/abs/astro-ph/0402430}
  {astro-ph/0402430} \BibitemShut {NoStop}%
\bibitem [{\citenamefont {{Levinson}}\ and\ \citenamefont
  {{Cerutti}}(2018)}]{Levinson_2018}%
  \BibitemOpen
  \bibfield  {author} {\bibinfo {author} {\bibfnamefont {A.}~\bibnamefont
  {{Levinson}}}\ and\ \bibinfo {author} {\bibfnamefont {B.}~\bibnamefont
  {{Cerutti}}},\ }\href {\doibase 10.1051/0004-6361/201832915} {\bibfield
  {journal} {\bibinfo  {journal} {aap}\ }\textbf {\bibinfo {volume} {616}},\
  \bibinfo {eid} {A184} (\bibinfo {year} {2018})},\ \Eprint
  {http://arxiv.org/abs/1803.04427} {arXiv:1803.04427 [astro-ph.HE]}
  \BibitemShut {NoStop}%
\bibitem [{\citenamefont {{Chen}}\ and\ \citenamefont
  {{Yuan}}(2019)}]{Chen_2019}%
  \BibitemOpen
  \bibfield  {author} {\bibinfo {author} {\bibfnamefont {A.~Y.}\ \bibnamefont
  {{Chen}}}\ and\ \bibinfo {author} {\bibfnamefont {Y.}~\bibnamefont
  {{Yuan}}},\ }\href@noop {} {\bibfield  {journal} {\bibinfo  {journal} {arXiv
  e-prints}\ } (\bibinfo {year} {2019})},\ \Eprint
  {http://arxiv.org/abs/1908.06919} {arXiv:1908.06919 [astro-ph.HE]}
  \BibitemShut {NoStop}%
\bibitem [{\citenamefont {{Parfrey}}\ \emph {et~al.}(2019)\citenamefont
  {{Parfrey}}, \citenamefont {{Philippov}},\ and\ \citenamefont
  {{Cerutti}}}]{Parfrey_2019}%
  \BibitemOpen
  \bibfield  {author} {\bibinfo {author} {\bibfnamefont {K.}~\bibnamefont
  {{Parfrey}}}, \bibinfo {author} {\bibfnamefont {A.}~\bibnamefont
  {{Philippov}}}, \ and\ \bibinfo {author} {\bibfnamefont {B.}~\bibnamefont
  {{Cerutti}}},\ }\href {\doibase 10.1103/PhysRevLett.122.035101} {\bibfield
  {journal} {\bibinfo  {journal} {Physical Review Letters}\ }\textbf {\bibinfo
  {volume} {122}},\ \bibinfo {eid} {035101} (\bibinfo {year} {2019})},\ \Eprint
  {http://arxiv.org/abs/1810.03613} {arXiv:1810.03613 [astro-ph.HE]}
  \BibitemShut {NoStop}%
\bibitem [{\citenamefont {{Cerutti}}\ and\ \citenamefont
  {{Werner}}(2019)}]{Zeltron}%
  \BibitemOpen
  \bibfield  {author} {\bibinfo {author} {\bibfnamefont {B.}~\bibnamefont
  {{Cerutti}}}\ and\ \bibinfo {author} {\bibfnamefont {G.}~\bibnamefont
  {{Werner}}},\ }\href@noop {} {\enquote {\bibinfo {title} {{Zeltron: Explicit
  3D relativistic electromagnetic Particle-In-Cell code}},}\ } (\bibinfo {year}
  {2019}),\ \Eprint {http://arxiv.org/abs/1911.012} {ascl:1911.012}
  \BibitemShut {NoStop}%
\bibitem [{\citenamefont {{Cerutti}}\ \emph {et~al.}(2013)\citenamefont
  {{Cerutti}}, \citenamefont {{Werner}}, \citenamefont {{Uzdensky}},\ and\
  \citenamefont {{Begelman}}}]{Cerutti_2013}%
  \BibitemOpen
  \bibfield  {author} {\bibinfo {author} {\bibfnamefont {B.}~\bibnamefont
  {{Cerutti}}}, \bibinfo {author} {\bibfnamefont {G.~R.}\ \bibnamefont
  {{Werner}}}, \bibinfo {author} {\bibfnamefont {D.~A.}\ \bibnamefont
  {{Uzdensky}}}, \ and\ \bibinfo {author} {\bibfnamefont {M.~C.}\ \bibnamefont
  {{Begelman}}},\ }\href {\doibase 10.1088/0004-637X/770/2/147} {\bibfield
  {journal} {\bibinfo  {journal} {\apj}\ }\textbf {\bibinfo {volume} {770}},\
  \bibinfo {eid} {147} (\bibinfo {year} {2013})},\ \Eprint
  {http://arxiv.org/abs/1302.6247} {arXiv:1302.6247 [astro-ph.HE]} \BibitemShut
  {NoStop}%
\bibitem [{\citenamefont {{Cerutti}}\ \emph {et~al.}(2015)\citenamefont
  {{Cerutti}}, \citenamefont {{Philippov}}, \citenamefont {{Parfrey}},\ and\
  \citenamefont {{Spitkovsky}}}]{Cerutti_2015}%
  \BibitemOpen
  \bibfield  {author} {\bibinfo {author} {\bibfnamefont {B.}~\bibnamefont
  {{Cerutti}}}, \bibinfo {author} {\bibfnamefont {A.}~\bibnamefont
  {{Philippov}}}, \bibinfo {author} {\bibfnamefont {K.}~\bibnamefont
  {{Parfrey}}}, \ and\ \bibinfo {author} {\bibfnamefont {A.}~\bibnamefont
  {{Spitkovsky}}},\ }\href {\doibase 10.1093/mnras/stv042} {\bibfield
  {journal} {\bibinfo  {journal} {mnras}\ }\textbf {\bibinfo {volume} {448}},\
  \bibinfo {pages} {606} (\bibinfo {year} {2015})},\ \Eprint
  {http://arxiv.org/abs/1410.3757} {arXiv:1410.3757 [astro-ph.HE]} \BibitemShut
  {NoStop}%
\bibitem [{\citenamefont {Frolov}\ and\ \citenamefont
  {Novikov}(1998)}]{Novikov}%
  \BibitemOpen
  \bibfield  {author} {\bibinfo {author} {\bibfnamefont {V.~P.}\ \bibnamefont
  {Frolov}}\ and\ \bibinfo {author} {\bibfnamefont {I.~D.}\ \bibnamefont
  {Novikov}},\ }\href@noop {} {\emph {\bibinfo {title} {Black Hole Physics:
  Basic Concepts and New Developments}}},\ \bibinfo {edition} {1st}\ ed.\
  (\bibinfo  {publisher} {Springer},\ \bibinfo {year} {1998})\BibitemShut
  {NoStop}%
\bibitem [{\citenamefont {{Blumenthal}}\ and\ \citenamefont
  {{Gould}}(1970)}]{Blumenthal_1970}%
  \BibitemOpen
  \bibfield  {author} {\bibinfo {author} {\bibfnamefont {G.~R.}\ \bibnamefont
  {{Blumenthal}}}\ and\ \bibinfo {author} {\bibfnamefont {R.~J.}\ \bibnamefont
  {{Gould}}},\ }\href {\doibase 10.1103/RevModPhys.42.237} {\bibfield
  {journal} {\bibinfo  {journal} {Reviews of Modern Physics}\ }\textbf
  {\bibinfo {volume} {42}},\ \bibinfo {pages} {237} (\bibinfo {year}
  {1970})}\BibitemShut {NoStop}%
\bibitem [{\citenamefont {{Jones}}(1968)}]{Jones_1968}%
  \BibitemOpen
  \bibfield  {author} {\bibinfo {author} {\bibfnamefont {F.~C.}\ \bibnamefont
  {{Jones}}},\ }\href {\doibase 10.1103/PhysRev.167.1159} {\bibfield  {journal}
  {\bibinfo  {journal} {Physical Review}\ }\textbf {\bibinfo {volume} {167}},\
  \bibinfo {pages} {1159} (\bibinfo {year} {1968})}\BibitemShut {NoStop}%
\bibitem [{\citenamefont {{Gould}}\ and\ \citenamefont
  {{Schr{\'e}der}}(1967)}]{Gould_1967}%
  \BibitemOpen
  \bibfield  {author} {\bibinfo {author} {\bibfnamefont {R.~J.}\ \bibnamefont
  {{Gould}}}\ and\ \bibinfo {author} {\bibfnamefont {G.~P.}\ \bibnamefont
  {{Schr{\'e}der}}},\ }\href {\doibase 10.1103/PhysRev.155.1404} {\bibfield
  {journal} {\bibinfo  {journal} {Physical Review}\ }\textbf {\bibinfo {volume}
  {155}},\ \bibinfo {pages} {1404} (\bibinfo {year} {1967})}\BibitemShut
  {NoStop}%
\bibitem [{\citenamefont {{Bonometto}}\ and\ \citenamefont
  {{Rees}}(1971)}]{Bonometto_1971}%
  \BibitemOpen
  \bibfield  {author} {\bibinfo {author} {\bibfnamefont {S.}~\bibnamefont
  {{Bonometto}}}\ and\ \bibinfo {author} {\bibfnamefont {M.~J.}\ \bibnamefont
  {{Rees}}},\ }\href {\doibase 10.1093/mnras/152.1.21} {\bibfield  {journal}
  {\bibinfo  {journal} {mnras}\ }\textbf {\bibinfo {volume} {152}},\ \bibinfo
  {pages} {21} (\bibinfo {year} {1971})}\BibitemShut {NoStop}%
\bibitem [{\citenamefont {{Aharonian}}\ \emph {et~al.}(1983)\citenamefont
  {{Aharonian}}, \citenamefont {{Atoian}},\ and\ \citenamefont
  {{Nagapetian}}}]{Aharonian_1983}%
  \BibitemOpen
  \bibfield  {author} {\bibinfo {author} {\bibfnamefont {F.~A.}\ \bibnamefont
  {{Aharonian}}}, \bibinfo {author} {\bibfnamefont {A.~M.}\ \bibnamefont
  {{Atoian}}}, \ and\ \bibinfo {author} {\bibfnamefont {A.~M.}\ \bibnamefont
  {{Nagapetian}}},\ }\href@noop {} {\bibfield  {journal} {\bibinfo  {journal}
  {Astrofizika}\ }\textbf {\bibinfo {volume} {19}},\ \bibinfo {pages} {323}
  (\bibinfo {year} {1983})}\BibitemShut {NoStop}%
\bibitem [{\citenamefont {{Komissarov}}\ and\ \citenamefont
  {{McKinney}}(2007)}]{Komissarov_2007}%
  \BibitemOpen
  \bibfield  {author} {\bibinfo {author} {\bibfnamefont {S.~S.}\ \bibnamefont
  {{Komissarov}}}\ and\ \bibinfo {author} {\bibfnamefont {J.~C.}\ \bibnamefont
  {{McKinney}}},\ }\href {\doibase 10.1111/j.1745-3933.2007.00301.x} {\bibfield
   {journal} {\bibinfo  {journal} {mnras}\ }\textbf {\bibinfo {volume} {377}},\
  \bibinfo {pages} {L49} (\bibinfo {year} {2007})},\ \Eprint
  {http://arxiv.org/abs/astro-ph/0702269} {astro-ph/0702269} \BibitemShut
  {NoStop}%
\bibitem [{\citenamefont {{Neronov}}\ and\ \citenamefont
  {{Aharonian}}(2007)}]{Neronov_2007}%
  \BibitemOpen
  \bibfield  {author} {\bibinfo {author} {\bibfnamefont {A.}~\bibnamefont
  {{Neronov}}}\ and\ \bibinfo {author} {\bibfnamefont {F.~A.}\ \bibnamefont
  {{Aharonian}}},\ }\href {\doibase 10.1086/522199} {\bibfield  {journal}
  {\bibinfo  {journal} {\apj}\ }\textbf {\bibinfo {volume} {671}},\ \bibinfo
  {pages} {85} (\bibinfo {year} {2007})},\ \Eprint
  {http://arxiv.org/abs/0704.3282} {arXiv:0704.3282 [astro-ph]} \BibitemShut
  {NoStop}%
\bibitem [{\citenamefont {{Event Horizon Telescope Collaboration}}\ \emph
  {et~al.}(2019{\natexlab{b}})\citenamefont {{Event Horizon Telescope
  Collaboration}}, \citenamefont {{Akiyama}}, \citenamefont {{Alberdi}},
  \citenamefont {{Alef}}, \citenamefont {{Asada}}, \citenamefont {{Azulay}},
  \citenamefont {{Baczko}}, \citenamefont {{Ball}}, \citenamefont
  {{Balokovi{\'c}}}, \citenamefont {{Barrett}} \emph {et~al.}}]{EHT_5}%
  \BibitemOpen
  \bibfield  {author} {\bibinfo {author} {\bibnamefont {{Event Horizon
  Telescope Collaboration}}}, \bibinfo {author} {\bibfnamefont
  {K.}~\bibnamefont {{Akiyama}}}, \bibinfo {author} {\bibfnamefont
  {A.}~\bibnamefont {{Alberdi}}}, \bibinfo {author} {\bibfnamefont
  {W.}~\bibnamefont {{Alef}}}, \bibinfo {author} {\bibfnamefont
  {K.}~\bibnamefont {{Asada}}}, \bibinfo {author} {\bibfnamefont
  {R.}~\bibnamefont {{Azulay}}}, \bibinfo {author} {\bibfnamefont {A.-K.}\
  \bibnamefont {{Baczko}}}, \bibinfo {author} {\bibfnamefont {D.}~\bibnamefont
  {{Ball}}}, \bibinfo {author} {\bibfnamefont {M.}~\bibnamefont
  {{Balokovi{\'c}}}}, \bibinfo {author} {\bibfnamefont {J.}~\bibnamefont
  {{Barrett}}},  \emph {et~al.},\ }\href {\doibase 10.3847/2041-8213/ab0f43}
  {\bibfield  {journal} {\bibinfo  {journal} {\apj}\ }\textbf {\bibinfo
  {volume} {875}},\ \bibinfo {eid} {L5} (\bibinfo {year}
  {2019}{\natexlab{b}})},\ \Eprint {http://arxiv.org/abs/1906.11242}
  {arXiv:1906.11242} \BibitemShut {NoStop}%
\bibitem [{\citenamefont {{Abdo}}\ \emph {et~al.}(2009)\citenamefont {{Abdo}},
  \citenamefont {{Ackermann}}, \citenamefont {{Ajello}}, \citenamefont
  {{Atwood}}, \citenamefont {{Axelsson}}, \citenamefont {{Baldini}},
  \citenamefont {{Ballet}}, \citenamefont {{Barbiellini}}, \citenamefont
  {{Bastieri}}, \citenamefont {{Bechtol}} \emph {et~al.}}]{Abdo_2009}%
  \BibitemOpen
  \bibfield  {author} {\bibinfo {author} {\bibfnamefont {A.~A.}\ \bibnamefont
  {{Abdo}}}, \bibinfo {author} {\bibfnamefont {M.}~\bibnamefont {{Ackermann}}},
  \bibinfo {author} {\bibfnamefont {M.}~\bibnamefont {{Ajello}}}, \bibinfo
  {author} {\bibfnamefont {W.~B.}\ \bibnamefont {{Atwood}}}, \bibinfo {author}
  {\bibfnamefont {M.}~\bibnamefont {{Axelsson}}}, \bibinfo {author}
  {\bibfnamefont {L.}~\bibnamefont {{Baldini}}}, \bibinfo {author}
  {\bibfnamefont {J.}~\bibnamefont {{Ballet}}}, \bibinfo {author}
  {\bibfnamefont {G.}~\bibnamefont {{Barbiellini}}}, \bibinfo {author}
  {\bibfnamefont {D.}~\bibnamefont {{Bastieri}}}, \bibinfo {author}
  {\bibfnamefont {K.}~\bibnamefont {{Bechtol}}},  \emph {et~al.},\ }\href
  {\doibase 10.1088/0004-637X/707/1/55} {\bibfield  {journal} {\bibinfo
  {journal} {\apj}\ }\textbf {\bibinfo {volume} {707}},\ \bibinfo {pages} {55}
  (\bibinfo {year} {2009})},\ \Eprint {http://arxiv.org/abs/0910.3565}
  {arXiv:0910.3565 [astro-ph.HE]} \BibitemShut {NoStop}%
\bibitem [{\citenamefont {{Mo{\'s}cibrodzka}}\ \emph
  {et~al.}(2011)\citenamefont {{Mo{\'s}cibrodzka}}, \citenamefont {{Gammie}},
  \citenamefont {{Dolence}},\ and\ \citenamefont
  {{Shiokawa}}}]{Moscibrodzka_2011}%
  \BibitemOpen
  \bibfield  {author} {\bibinfo {author} {\bibfnamefont {M.}~\bibnamefont
  {{Mo{\'s}cibrodzka}}}, \bibinfo {author} {\bibfnamefont {C.~F.}\ \bibnamefont
  {{Gammie}}}, \bibinfo {author} {\bibfnamefont {J.~C.}\ \bibnamefont
  {{Dolence}}}, \ and\ \bibinfo {author} {\bibfnamefont {H.}~\bibnamefont
  {{Shiokawa}}},\ }\href {\doibase 10.1088/0004-637X/735/1/9} {\bibfield
  {journal} {\bibinfo  {journal} {\apj}\ }\textbf {\bibinfo {volume} {735}},\
  \bibinfo {eid} {9} (\bibinfo {year} {2011})},\ \Eprint
  {http://arxiv.org/abs/1104.2042} {arXiv:1104.2042 [astro-ph.HE]} \BibitemShut
  {NoStop}%
\bibitem [{\citenamefont {{Goldreich}}\ and\ \citenamefont
  {{Julian}}(1969)}]{Goldreich_1969}%
  \BibitemOpen
  \bibfield  {author} {\bibinfo {author} {\bibfnamefont {P.}~\bibnamefont
  {{Goldreich}}}\ and\ \bibinfo {author} {\bibfnamefont {W.~H.}\ \bibnamefont
  {{Julian}}},\ }\href {\doibase 10.1086/150119} {\bibfield  {journal}
  {\bibinfo  {journal} {\apj}\ }\textbf {\bibinfo {volume} {157}},\ \bibinfo
  {pages} {869} (\bibinfo {year} {1969})}\BibitemShut {NoStop}%
\bibitem [{\citenamefont {{Timokhin}}\ and\ \citenamefont
  {{Harding}}(2019)}]{Timokhin_2019}%
  \BibitemOpen
  \bibfield  {author} {\bibinfo {author} {\bibfnamefont {A.~N.}\ \bibnamefont
  {{Timokhin}}}\ and\ \bibinfo {author} {\bibfnamefont {A.~K.}\ \bibnamefont
  {{Harding}}},\ }\href {\doibase 10.3847/1538-4357/aaf050} {\bibfield
  {journal} {\bibinfo  {journal} {\apj}\ }\textbf {\bibinfo {volume} {871}},\
  \bibinfo {eid} {12} (\bibinfo {year} {2019})},\ \Eprint
  {http://arxiv.org/abs/1803.08924} {arXiv:1803.08924 [astro-ph.HE]}
  \BibitemShut {NoStop}%
\bibitem [{\citenamefont
  {{Komissarov}}(2004{\natexlab{b}})}]{Komissarov_2004a}%
  \BibitemOpen
  \bibfield  {author} {\bibinfo {author} {\bibfnamefont {S.~S.}\ \bibnamefont
  {{Komissarov}}},\ }\href {\doibase 10.1111/j.1365-2966.2004.07598.x}
  {\bibfield  {journal} {\bibinfo  {journal} {mnras}\ }\textbf {\bibinfo
  {volume} {350}},\ \bibinfo {pages} {427} (\bibinfo {year}
  {2004}{\natexlab{b}})}\BibitemShut {NoStop}%
\bibitem [{\citenamefont {{Nathanail}}\ and\ \citenamefont
  {{Contopoulos}}(2014)}]{Nathanail_2014}%
  \BibitemOpen
  \bibfield  {author} {\bibinfo {author} {\bibfnamefont {A.}~\bibnamefont
  {{Nathanail}}}\ and\ \bibinfo {author} {\bibfnamefont {I.}~\bibnamefont
  {{Contopoulos}}},\ }\href {\doibase 10.1088/0004-637X/788/2/186} {\bibfield
  {journal} {\bibinfo  {journal} {\apj}\ }\textbf {\bibinfo {volume} {788}},\
  \bibinfo {eid} {186} (\bibinfo {year} {2014})},\ \Eprint
  {http://arxiv.org/abs/1404.0549} {arXiv:1404.0549 [astro-ph.HE]} \BibitemShut
  {NoStop}%
\bibitem [{\citenamefont {{Takahashi}}\ \emph {et~al.}(1990)\citenamefont
  {{Takahashi}}, \citenamefont {{Nitta}}, \citenamefont {{Tatematsu}},\ and\
  \citenamefont {{Tomimatsu}}}]{Takahashi_1990}%
  \BibitemOpen
  \bibfield  {author} {\bibinfo {author} {\bibfnamefont {M.}~\bibnamefont
  {{Takahashi}}}, \bibinfo {author} {\bibfnamefont {S.}~\bibnamefont
  {{Nitta}}}, \bibinfo {author} {\bibfnamefont {Y.}~\bibnamefont
  {{Tatematsu}}}, \ and\ \bibinfo {author} {\bibfnamefont {A.}~\bibnamefont
  {{Tomimatsu}}},\ }\href {\doibase 10.1086/169331} {\bibfield  {journal}
  {\bibinfo  {journal} {\apj}\ }\textbf {\bibinfo {volume} {363}},\ \bibinfo
  {pages} {206} (\bibinfo {year} {1990})}\BibitemShut {NoStop}%
\bibitem [{\citenamefont {{Tchekhovskoy}}\ \emph {et~al.}(2010)\citenamefont
  {{Tchekhovskoy}}, \citenamefont {{Narayan}},\ and\ \citenamefont
  {{McKinney}}}]{Tchekhovskoy_2010}%
  \BibitemOpen
  \bibfield  {author} {\bibinfo {author} {\bibfnamefont {A.}~\bibnamefont
  {{Tchekhovskoy}}}, \bibinfo {author} {\bibfnamefont {R.}~\bibnamefont
  {{Narayan}}}, \ and\ \bibinfo {author} {\bibfnamefont {J.~C.}\ \bibnamefont
  {{McKinney}}},\ }\href {\doibase 10.1088/0004-637X/711/1/50} {\bibfield
  {journal} {\bibinfo  {journal} {\apj}\ }\textbf {\bibinfo {volume} {711}},\
  \bibinfo {pages} {50} (\bibinfo {year} {2010})},\ \Eprint
  {http://arxiv.org/abs/0911.2228} {arXiv:0911.2228 [astro-ph.HE]} \BibitemShut
  {NoStop}%
\end{thebibliography}%



\end{document}